\documentclass[times,twocolumn,final]{elsarticle}

\usepackage{cag}
\usepackage{framed,multirow}

\usepackage{amssymb}
\usepackage{latexsym}

\usepackage{url}
\usepackage{xcolor}
\definecolor{newcolor}{rgb}{.8,.349,.1}

\usepackage[normalem]{ulem}
\usepackage{hyperref}

\hypersetup{
    colorlinks=true,
    linkcolor=blue,
    filecolor=blue,      
    urlcolor=blue,
    citecolor=blue
}

\usepackage[switch,pagewise]{lineno} 

\usepackage{siunitx}
\usepackage{orcidlink}

\journal{Computers \& Graphics}

\begin{document}

\verso{Open Access: \url{https://doi.org/10.1016/j.cag.2026.104727}}

\begin{frontmatter}

    \title{MIA: A Visual Analytics System for Multimodal Spectral Imaging Data}%

    \author[1]{Hennes Rave \orcidlink{0000-0001-8662-3767}\corref{correspondingAuthor}}
    \author[2,3]{Katharina Kronenberg \orcidlink{0009-0002-0814-0043}}
    \author[2]{Hannes G{\"o}dde \orcidlink{0009-0000-9573-3802}}
    \author[2]{Lea Tobergte \orcidlink{0009-0005-9320-5623}}
    \author[2]{Michael Holtkamp}
    \author[4]{Julia Werner}
    \author[4]{Peter Bohrer}
    \author[4]{Fabian Loh{\"o}fer}
    \author[5]{Rickmer Braren \orcidlink{0000-0001-6039-6957}}
    \author[3]{David Clases \orcidlink{0000-0003-3880-9385}}
    \author[2]{Uwe Karst \orcidlink{0000-0002-1774-6787}}
    \author[1]{Lars Linsen \orcidlink{0000-0002-6168-8748}}

    \address[1]{Computer Science Department, University of M{\"u}nster, 48149 M{\"u}nster, Germany}
    \address[2]{Institute of Inorganic and Analytical Chemistry, University of M{\"u}nster, 48149, M{\"u}nster, Germany}
    \address[3]{NanoMicroLab, Institute of Chemistry, University of Graz, 8010 Graz, Austria}
    \address[4]{Institute of Diagnostic and Interventional Radiology, School of Medicine and Health, Technical University of Munich, 81675 Munich, Germany}
    \address[5]{Department of Diagnostic and Interventional Radiology and Nuclear Medicine, University Medical Center Hamburg-Eppendorf, 20246 Hamburg, Germany}

    \cortext[correspondingAuthor]{Corresponding author: hennes.rave@uni-muenster.de}
    \emailauthor{hennes.rave@uni-muenster.de}{Hennes Rave}

    \accepted{in Computers \& Graphics (VCBM 2026)}
    \availableonline{at \url{https://doi.org/10.1016/j.cag.2026.104727}}

    \begin{abstract}
        Hyperspectral bioimaging techniques such as infrared~(IR) microscopy and laser ablation-inductively coupled plasma-mass spectrometry (LA-ICP-MS) produce high-dimensional, spatially resolved datasets that require sophisticated analysis to reveal chemically and anatomically meaningful structures.
        Existing software solutions are typically modality-specific and cover only parts of the analytical workflow, forcing researchers to transfer data across multiple tools and manually reconcile results.
        We present MIA (Multiscale Image Analysis), a modality-agnostic visual analysis environment for analytical chemistry that integrates the full exploratory workflow -- from spectral preprocessing and dimensionality reduction to interactive segmentation and spectral similarity analysis -- within a single, tightly coupled interface.
        MIA supports hierarchical and landmark-based embeddings to handle datasets of varying scale and complexity, interactive and automatic segmentation with a shared state across all linked views, and multimodal analysis of co-registered datasets from different instruments.
        We demonstrate the effectiveness of MIA through three use cases drawn from real analytical chemistry workflows: (1)~the recovery of biologically meaningful tissue compartments through derivative preprocessing and hierarchical embedding, (2)~pigment identification via spectral similarity search with spatial overview, and (3)~multimodal tissue characterization combining molecular IR and elemental LA-ICP-MS data.
        Qualitative feedback from domain expert collaborators confirms that MIA reduces the need for tool-switching and supports analytical insights that are difficult to obtain with existing software.
    \end{abstract}


\end{frontmatter}


\section{Introduction}

Hyperspectral bioimaging has become an increasingly important tool across the biomedical and analytical sciences. Techniques such as quantum cascade laser~(QCL)-based infrared~(IR) microscopy, micro x-ray fluorescence~(micro-XRF), and laser ablation-inductively coupled plasma-mass spectrometry~(LA-ICP-MS) acquire spatially resolved measurements with up to thousands of spectral channels, producing high-dimensional datasets that encode rich molecular and elemental information about biological tissue, materials, and environmental samples~\cite{doble_2021_la_icp_ms, kronenberg_2023_multimodal, kroger_2014_quantum, schwaighofer_2017_quantum}. The growing availability of these instruments and the emergence of workflows that combine multiple hyperspectral modalities on the same specimen have created a pressing need for flexible, interactive software that can support the full analytical pipeline.

Extracting meaningful information from hyperspectral bioimaging data is inherently a multi-step process. Raw spectra typically require preprocessing such as baseline correction, normalization, or the computation of spectral derivatives before structure becomes apparent. Dimensionality reduction is then used to embed the high-dimensional spectra into a low-dimensional space where spectrally similar pixels cluster together, providing a basis for segmentation. Segmentation itself is rarely fully automatic: domain experts routinely refine automatic clusterings by inspecting individual segments in both spatial and spectral terms, merging or splitting regions based on domain knowledge. Finally, the resulting segments are interpreted through spectral exploration, comparing per-segment summary spectra, computing similarities to reference spectra, or examining the distribution of individual channels across segments. Crucially, these steps are not sequential but iterative: a segmentation may prompt re-examination of the embedding, which may in turn motivate a different preprocessing strategy. Manual segmentation by domain experts is, however, inherently subjective, with variability between observers depending on training and institutional conventions~\cite{walker2006_quantification}. Reproducible algorithmic approaches that integrate expert refinement with transparent automatic clustering can mitigate this variability.

Existing software solutions address parts of this workflow, but rarely the whole. Tools designed for IR spectroscopy, such as Quasar~\cite{toplak_2017_quasar, toplak_2021_quasar}, provide strong preprocessing pipelines but follow a node-based workflow model that limits interactive exploration and lacks support for hierarchical or scalable embedding methods. Tools targeting LA-ICP-MS focus primarily on elemental map generation and quantitative analysis, with limited support for embedding-based exploration or interactive segmentation across linked views. Broader mass spectrometry imaging platforms such as Cardinal~\cite{bemis_2015_cardinal, bemis_2023_cardinal} and commercial tools such as SCiLS Lab provide statistical analysis capabilities but are typically instrument-specific and restrict cross-modality workflows to vendor-specific datasets. As a result, researchers routinely transfer data between multiple disconnected tools, a process that is time-consuming, error-prone, and difficult to reproduce.

We present \textit{MIA} (Multiscale Image Analysis), a modality-agnostic visual analysis environment designed to support the full exploratory workflow for hyperspectral bioimaging data within a single, tightly coupled interface. MIA treats any dataset as a collection of pixels with associated spectra and spatial coordinates. Because this data model makes no modality-specific assumptions, the same interface and workflow apply to any modality that can be represented in this form. We demonstrate this on IR microscopy, LA-ICP-MS, and micro-XRF, as well as their multimodal combinations. A shared segmentation state propagates across all linked views, ensuring that every interaction, whether a lasso selection in the embedding, a manual edit in the image viewer, or an automatic clustering, is immediately reflected throughout the interface. To handle the scale and complexity of real-world datasets, MIA supports landmark-based UMAP~\cite{mcinnes_2018_umap} for datasets of millions of pixels, as well as hierarchical embeddings that allow users to iteratively refine the analysis within individual segments. MIA is open source and freely available at \url{https://github.com/hennesrave/multiscale-image-analysis}.

The contributions of this paper are:
\begin{itemize}
    \item A modality-agnostic visual analysis environment for (multimodal) hyperspectral bioimaging that integrates spectral preprocessing, embedding-based dimensionality reduction, interactive segmentation, and spectral exploration within a single linked-view interface with a shared segmentation state.
    \item The integration of interactive, user-driven re-embedding of selected segments into this linked-view workflow, allowing tissue structure to be resolved at multiple scales within a single session. Rather than proposing a new hierarchical dimensionality reduction algorithm, we contribute the coupling of recursive re-embedding with the shared segmentation state and spectral preprocessing, so that subclusters can be inspected, refined, and validated spatially and spectrally in place.
    \item A validation of MIA through three real-world analytical chemistry workflows, demonstrating that the integrated design reduces tool-switching, supports emergent analytical strategies beyond those explicitly designed for, and enables insights that neither individual modalities nor manual approaches
          can reliably produce.
\end{itemize}

In the following, we first review related work on hyperspectral imaging software and embedding-based analysis methods in Section~\ref{sec:related}. We, then, present the MIA system, its data model, its visual encodings, and interaction mechanisms in Section~\ref{sec:system}. We apply MIA within three use cases drawn from real analytical chemistry workflows in Section~\ref{sec:usecases} and discuss expert feedback, limitations, and directions for future work in Section~\ref{sec:discussion}.

\section{Related Work}
\label{sec:related}

\subsection{IR Spectroscopy Imaging Software}

Infrared microscopy produces spatially resolved hyperspectral data that requires dedicated preprocessing before meaningful analysis is possible. The most widely used open-source tool for this domain is Quasar~\cite{toplak_2017_quasar, toplak_2021_quasar}, a distribution of the Orange data-mining platform~\cite{demsar_2013_orange} extended with spectroscopy-specific components. Quasar supports a broad range of preprocessing operations and integrates machine learning methods such as PCA~\cite{jolliffe_2005_principal}, clustering, and classification into a visual workflow-builder approach, in which components are connected as nodes in a pipeline. While this approach is flexible and accessible, it differs fundamentally from MIA's design: Quasar does not provide linked spatial and spectral views that share a common segmentation state, does not support hierarchical or landmark-based UMAP for interactive exploration of large datasets, and is not designed for multimodal analysis across different instrument types. Commercial tools such as Cytospec~\cite{cytospec_spectroscopy_2026} and Bruker OPUS~\cite{bruker_opus_2026} offer IR-specific preprocessing and visualization but are hardware-tied and offer limited support for modern dimensionality reduction or interactive segmentation.

\subsection{LA-ICP-MS Software}


LA-ICP-MS bioimaging software has historically focused on data reduction and elemental map generation. Open-source tools such as LA-iMageS~\cite{lopez_fernandez_2016_la_images} provide automated processing and visualization of elemental distribution images, while commercial platforms such as iolite from Elemental Scientific Lasers~\cite{paton_2011_iolite} and HDIP from Teledyne Photon Machines~\cite{teledyne_hdip} offer comprehensive data reduction pipelines, automated workflows, and integrated 2D and 3D visualization of elemental maps. HDIP additionally includes feature recognition and automated image processing tools. Several other tools targeting LA-ICP-MS data exist, including Pew\textsuperscript{2}~\cite{lockwood2021_pewpew}, an open-source image processing package compatible with common ICP-MS vendors, MeXpose~\cite{braun2024_mexpose}, which provides quantitative single-cell metal bioaccumulation analysis, and TOFHunter~\cite{andrews2025_tofhunter}, which enables rapid untargeted screening of ICP-TOF-MS data. While these platforms differ in focus and feature set, none provide an integrated workflow that combines embedding-based dimensionality reduction, interactive segmentation across linked spatial and spectral views, and modality-agnostic multimodal analysis as in MIA. The growing application of UMAP-based analysis to LA-ICP-TOFMS datasets~\cite{kronenberg_2025_umap,umfahrer2026_mineral} highlights the scientific need for embedding-based exploration of this data type.

\subsection{Molecular Mass Spectrometry Imaging Software}

Molecular mass spectrometry imaging (MSI) is a related field that has attracted more software development effort, providing useful context for MIA's positioning. Cardinal~\cite{bemis_2015_cardinal, bemis_2023_cardinal} is an R-based open-source package offering comprehensive statistical analysis of MSI data, including preprocessing, spatial segmentation, and classification. While analytically powerful, Cardinal operates primarily as a scripting environment and provides limited support for interactive visual exploration. The commercial SCiLS Lab platform offers a more interactive interface but is instrument-specific. M2aia~\cite{cordes_2021_m2aia} is an open-source application providing interactive visualization and 3D reconstruction for multimodal MSI datasets, and is the closest existing tool to MIA in terms of interactivity and multimodal support. However, M2aia is designed specifically for molecular MS modalities and does not support IR microscopy or LA-ICP-MS data. More broadly, as noted in recent work, most MSI tools target specific steps in the analytical pipeline or specific instrument families, leaving researchers to integrate multiple tools for a complete workflow. MIA addresses this gap across the IR, LA-ICP-MS, and micro-XRF domains by providing a unified, modality-agnostic environment that spans the full workflow from preprocessing to spectral exploration.

\subsection{Dimensionality Reduction and Embedding in Bioimaging}

Dimensionality reduction has become a central component of exploratory analysis for high-dimensional bioimaging data. Uniform Manifold Approximation and Projection~(UMAP)~\cite{mcinnes_2018_umap} has emerged as the method of choice for this task due to its ability to preserve both local and global structure at superior runtime compared to t-SNE~\cite{maaten_2008_tsne}, making it particularly suited to the large datasets produced by modern imaging instruments. A recent survey on the application of dimensionality reduction in different application domains, including biology and chemistry, confirms this observation~\cite{10992274}. UMAP has been applied extensively in single-cell transcriptomics and, more recently, in spatial bioimaging contexts~\cite{kronenberg_2023_multimodal,kronenberg_2025_umap}, where it enables the identification of spatially coherent tissue regions that are not apparent in individual channel images. For very large datasets, landmark-based UMAP approximations reduce computational cost by computing the embedding on a representative subset of points and projecting the remaining data, making interactive exploration of datasets with millions of pixels feasible. For spatially or spectrally complex data, hierarchical approaches that iteratively re-embed subsets of the data allow users to refine the analysis within individual clusters, revealing fine-grained structure that global embeddings may obscure~\cite{https://doi.org/10.1111/cgf.12878,8f6fd76ed94d4bd38316929d229407d3,kronenberg_2025_umap}. Dedicated hierarchical dimensionality reduction methods construct such multiscale representations algorithmically: Hierarchical Stochastic Neighbor Embedding~(HSNE)~\cite{https://doi.org/10.1111/cgf.12878} builds a hierarchy of non-linear similarities that can be explored from a coarse overview down to single data points, HUMAP~\cite{marcilio_2024_humap} extends this ``overview-first, details-on-demand'' strategy to UMAP, and Multiscale PHATE~\cite{kuchroo_2022_multiscale_phate} derives a continuum of granularities through a diffusion-condensation process. Such methods have also been applied directly to spectral imaging data: Abdelmoula et al.~\cite{abdelmoula_2018_hsne_msi} use HSNE to interactively explore 3D mass spectrometry imaging data at full spectral and spatial resolution, revealing spatiomolecular structures across scales. MIA does not implement a dedicated hierarchical embedding algorithm. Instead, it supports interactive, user-driven re-embedding of pixels belonging to selected segments, with the resulting subclusters folded back into the shared segmentation state for spatial and spectral inspection. These methods construct the multiscale hierarchy algorithmically and expose it for interactive slicing, so that a user-selected subset can be re-embedded on the fly from the corresponding level of the hierarchy. MIA differs not in whether the user can control which subset is refined, but in that it does not compute a dedicated hierarchical embedding at all. Its contribution is instead the tight coupling of user-driven re-embedding with the shared segmentation state and spectral preprocessing. Graph-based clustering methods such as the Leiden algorithm \cite{traag_2019_leiden} complement UMAP by operating directly on the nearest-neighbor graph used to construct the embedding, often producing more stable and well-connected clusters than $K$-means or density-based alternatives~\cite{wolf_2019_paga}. Such graph-based clustering can itself be made hierarchical: SCHNEL~\cite{abdelaal_2020_schnel}, for instance, applies community detection to the hierarchical neighborhood graph of HSNE, scaling clustering to millions of data points. MIA integrates all of these capabilities -- standard UMAP, a landmark-based approximation for large datasets, and interactive hierarchical re-embedding, combined with Leiden, HDBSCAN~\cite{mcinnes_2017_hdbscan}, and $K$-means clustering -- into a single interactive environment, directly connecting the embedding to spatial and spectral views through a shared segmentation state.

\subsection{Visual Analysis of Hyperspectral Imaging Data}
From the visualization community, Gerbil~\cite{jordan_2016_gerbil} is one of the most closely related general-purpose frameworks for interactive hyperspectral image analysis. It combines linked spectral and spatial views, interactive segmentation, and nonlinear false-color coding within a unified interface and has been applied to remote sensing and cultural heritage data. When compared to MIA, Gerbil is designed for reflectance spectroscopy rather than bioimaging modalities, does not support embedding-based dimensionality reduction or multimodal datasets, and targets a different user community. Gerbil's false-color coding is one instance of a more general idea, namely mapping low-dimensional embedding coordinates to color in order to recolor an image. This idea is implemented within the spatially mapped t-SNE of mass spectrometry imaging data by Abdelmoula et al.~\cite{abdelmoula_2016_tsne}, which projects spectra into a low-dimensional embedding and maps the embedding coordinates back onto the spatial domain as color. It is also implemented in the work of Evers et al.~\cite{evers_2021_uncertainty}, which maps a 3D embedding of spatially resolved correlations onto a 3D color space. MIA builds on this principle for its false-coloring, additionally applying a density-equalizing transformation to improve color contrast in dense embedding regions. 

In the cultural-heritage domain, Popa et al.~\cite{popa_2022_ris} present an interactive visual approach for endmember selection in reflectance imaging spectroscopy data, a hyperspectral modality related to those targeted by Gerbil. Jawad et al.~\cite{jawad_2020_interactive} propose an interactive visual analysis approach for molecular MSI data using linear and non-linear embeddings in coordinated image-space and spectral-space views, demonstrating that embedding-driven cluster selection can faithfully recover ground-truth image regions. Their approach is conceptually close to MIA's embedding-segmentation workflow, but targets molecular MSI specifically and does not support hierarchical or landmark-based embeddings, multimodal datasets, or interactive spectral similarity analysis. Several visual analysis systems from the visualization community target spectral data in other domains. Most closely related in spirit, InSpectr by Amirkhanov et al.~\cite{amirkhanov_2014_inspectr} integrates high-resolution X-ray computed tomography with low-resolution X-ray fluorescence spectral data for the multimodal analysis of industrial components. It combines linked spatial and spectral views, spectral transfer functions, and glyph-based composition overviews. In the clinical domain, Nunes et al.~\cite{nunes_2014_integrated} fuse magnetic resonance spectroscopy (MRS) with multi-modal radiology imaging to support tumor target-volume definition, and SpectraMosaic by Garrison et al.~\cite{garrison_2020_metabolite} enables the interactive exploration of metabolite ratios across cohorts in MRS studies. These systems share MIA's emphasis on linked, interactive analysis of spectral data, but target fundamentally different modalities -- industrial XRF and clinical MRS -- with far fewer spectral components than hyperspectral bioimaging data. In addition, they do not combine embedding-based dimensionality reduction with interactive segmentation across co-registered modalities. Further visualization-community contributions have addressed related subproblems. Zhu et al.~\cite{zhu_2021_interactive} use a cascade neural network to generate user-guided scatterplot embeddings for interactive hyperspectral image classification, with reuse of the trained network across time-varying datasets. Yu and Li~\cite{yu_2025_feature} propose a feature-selection-oriented interactive visual analysis approach that guides the selection of spectral bands through information-theoretic indicators. Both target remote sensing applications rather than bioimaging. At the opposite end of the modality spectrum, Cell2Cell by Mörth et al.~\cite{morth_2025_cell2cell} presents a visual analytics system for analyzing cell-cell interactions in 3D multi-channel tissue data, combining graph-based interaction analysis with cell-centered polarization views. While the multi-channel imaging context is related, Cell2Cell operates on volumetric protein-expression data with only a handful of channels, rather than on hyperspectral data with hundreds or thousands of channels per pixel. Cell2Cell is part of a broader body of visualization research on high-dimensional but non-hyperspectral tissue imaging, particularly in the spatial omics domain. ImaCytE by Somarakis et al.~\cite{somarakis_2021_imacyte} supports the visual exploration of cellular micro-environments in imaging mass cytometry data. Its companion system SpaCeCo~\cite{somarakis_2021_cohort} enables visual cohort comparison of spatially resolved single-cell data across specimens. Scope2Screen by Jessup et al.~\cite{jessup_2022_scope2screen} provides focus-and-context techniques for the assessment of highly multiplexed tissue images, and Facetto by Krueger et al.~\cite{krueger_2020_facetto} combines unsupervised and supervised learning for hierarchical phenotype analysis in multi-channel microscopy. Like Cell2Cell, these systems operate on data with at most a few dozen targeted, pre-selected markers per pixel, rather than the hundreds to thousands of largely untargeted spectral channels that characterize the hyperspectral modalities MIA addresses. This difference is consequential: hyperspectral data require spectral preprocessing and embedding to make latent chemical structure apparent, whereas marker-based imaging begins from interpretable, individually meaningful channels. Most closely related to MIA's interaction model, SEAL by Warchol et al.~\cite{warchol_2026_seal} links 2D embeddings of high-dimensional spatial imaging data with their spatial and morphological context. It allows selections to be defined and compared in both the embedding and image views and explains embedding structure through feature-importance scores. While this linked embedding-image selection model closely mirrors MIA's, SEAL targets morphology-rich imaging -- such as multiplexed tissue, satellite, and astronomical data -- and emphasizes explainability of the embedding, rather than the spectral preprocessing and multimodal integration central to MIA.

\section{Design Goals}
\label{sec:goals}

MIA was developed in close collaboration with researchers in analytical chemistry and bioimaging through an iterative process. Initial informal discussions with domain collaborators identified recurring pain points in existing workflows: the need to transfer data and intermediate results across multiple disconnected tools, the absence of interactive linked views maintaining a consistent segmentation, and the lack of support for combining data from different imaging modalities. Building on these observations and the gap identified in Section~\ref{sec:related}, we formulated the following design goals:

\textbf{G1~-- Integrated workflow.} Support the full analytical pipeline -- spectral preprocessing, dimensionality reduction, interactive segmentation, and spectral exploration -- within a single, tightly coupled environment, eliminating the need to transfer intermediate results between tools.

\textbf{G2~-- Shared segmentation state.} Maintain a single segmentation that propagates across all linked views, so that any interaction -- a lasso selection, an automatic clustering, or a manual edit -- is immediately and consistently reflected throughout the interface.

\textbf{G3~-- Modality-agnosticism.} Accommodate hyperspectral datasets from different imaging instruments under a common data model, enabling analysis of IR microscopy, LA-ICP-MS, micro-XRF, and multimodal combinations thereof without modification to the interface or workflow.

\textbf{G4~-- Scalability.} Handle datasets produced by modern imaging instruments -- including those with millions of pixels and hundreds of spectral channels -- without requiring specialized hardware.

\textbf{G5~-- Iterative exploration.} Support non-linear, iterative workflows in which users revisit earlier analysis steps -- such as adjusting preprocessing before recomputing an embedding -- within a single session. This is supported by analysis artifacts and application state being importable and exportable, so that earlier embeddings and segmentations can be stored, reloaded, and compared.

Achieving G3 posed specific design challenges: imaging modalities differ substantially in channel count (from a handful of elemental channels in LA-ICP-MS to several hundred spectral bands in IR microscopy), in pixel count and spatial resolution, and in the units and semantics of the spectral axis. MIA addresses these by treating every dataset uniformly as a collection of pixels, each associated with a spectrum of arbitrary length, with no assumptions about the nature or units of the spectral axis. When co-registered datasets from different instruments are combined, differences in spatial resolution are handled through user-guided alignment and resampling, and the risk of high-channel-count modalities dominating a joint embedding is mitigated through explicit modality weighting (Section~\ref{sec:system}).

The intended users of MIA are analytical chemists and bioimaging researchers who regularly acquire hyperspectral imaging data but do not have a strong computational background. These users currently rely on graphical tools -- instrument-specific vendor software or ad-hoc combinations of analysis packages -- and are not expected to write analysis scripts for exploratory tasks.

\section{MIA: System Overview}
\label{sec:system}

\begin{figure*}[t]
    \centering
    \includegraphics[width=\linewidth]{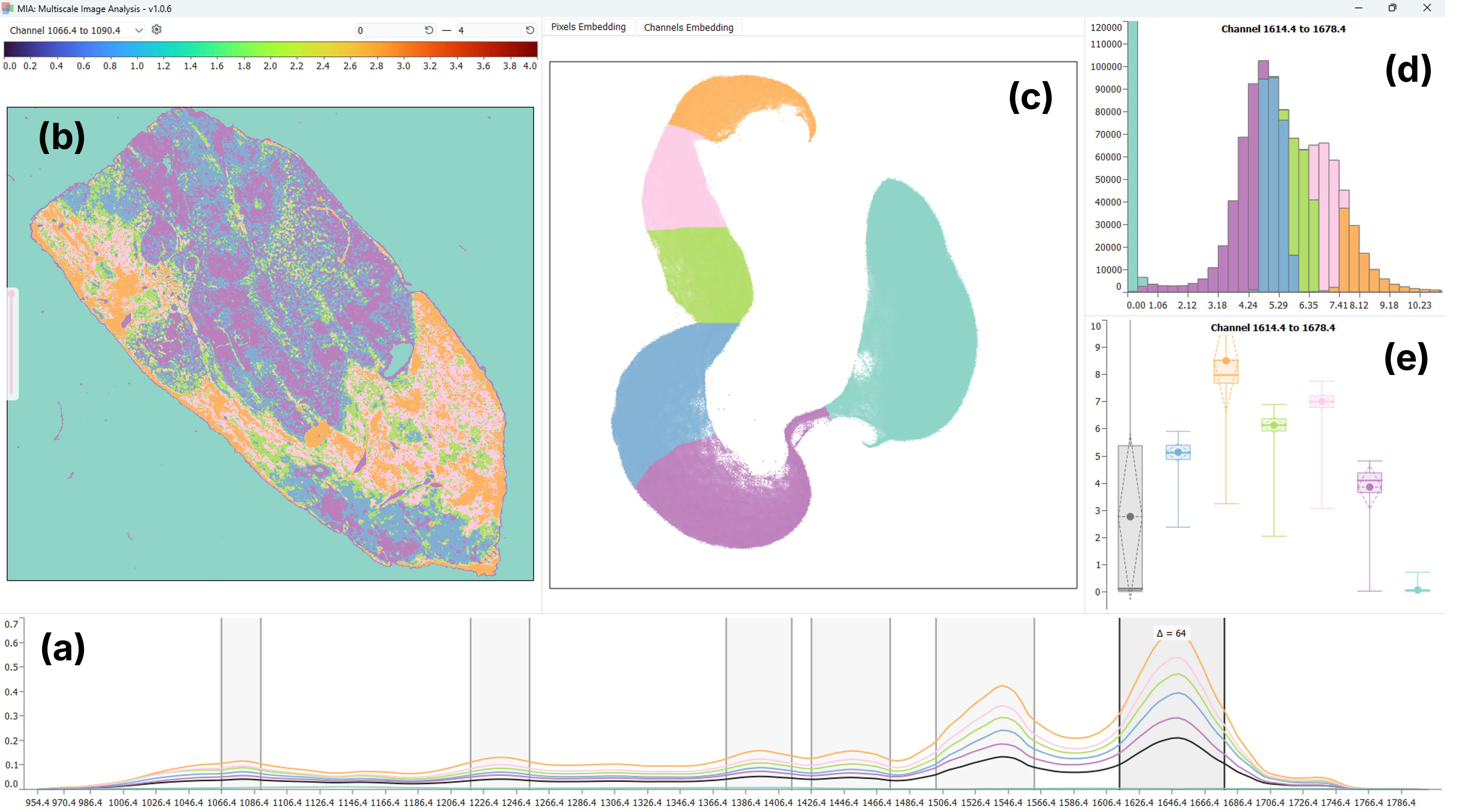}
    \caption{The MIA application window showing all five views on an IR microscopy dataset of a tumor-bearing rat liver~\cite{kronenberg_2023_multimodal}. (a)~Spectrum Viewer displaying per-segment average spectra and the global average in black; (b)~Image Viewer showing the segmentation overlay; (c)~Embedding Viewer (and Channel Glyph Viewer, accessible via tabs), showing a UMAP embedding with points color-coded by segment; (d)~Histogram Viewer showing the frequency distribution and (e)~Boxplot Viewer showing the intensity distribution for a selected feature.}
    \label{fig:teaser}
\end{figure*}

MIA is a desktop application for the interactive visual analysis of hyperspectral bioimaging data. Figure~\ref{fig:teaser} shows the main application window, which consists of five coordinated views: the Spectrum Viewer~(a), Image Viewer~(b), Embedding Viewer and Channel Glyph Viewer~(c, shared via tabs), and Histogram/Boxplot Viewers~(d,~e). The views are linked through a shared segmentation state~(G2): any change to the segmentation, whether through manual lasso selection, automatic clustering, or segment merging, is immediately reflected in all views simultaneously. Most functionality is accessible through context menus, keeping the interface uncluttered while supporting a wide range of operations. MIA is modality-agnostic~(G3): it operates on a common data model that accommodates IR microscopy, LA-ICP-MS, micro-XRF, and other hyperspectral imaging modalities without modification. MIA is implemented in C++ and is openly available at \url{https://github.com/hennesrave/multiscale-image-analysis}.

\subsection{Data Model and Terminology}

MIA operates on a \textit{dataset} consisting of \textit{pixels}, i.e., individual measurement points, each associated with a \textit{spectrum} (one intensity value per channel) and a pair of spatial coordinates. The dataset is organized as a matrix of $N$ pixels by $C$ channels, where $C$ ranges from tens of channels for
modalities such as LA-ICP-MS to several thousand for micro-XRF. Within a single dataset, the channels are ordered along the spectral axis (e.g., wavenumber for IR microscopy or mass-to-charge ratio for MS) and share a common physical unit and, typically, a regular sampling interval. This shared axis is what makes channel-wise operations, such as derivatives, well-defined. Across modalities, the units and ranges differ, which is why intensities are normalized before they are combined in a joint embedding (Section~\ref{sec:multimodal}).

A \textit{segmentation} partitions the set of pixels into disjoint \textit{segments}, each with an assigned name and color. A segment may consist of multiple disjoint regions in the image, reflecting the fact that pixels are typically assigned to segments based on spectral similarity rather than spatial proximity. The segmentation is global: it is shared across all views and all modalities in a multimodal session, ensuring consistency throughout the analysis. Pixels not yet assigned to any segment are treated as belonging to an implicit background.

A \textit{feature} maps each pixel to a single scalar value and is used as the basis for image display, histograms, and boxplots. Features can be defined by selecting a single channel, integrating a range of channels with optional baseline correction, or combining existing features through arithmetic operations.

An \textit{embedding} maps pixels -- or, in the channel embedding, channels -- to 2D points such that similar items are positioned close together. Embeddings are computed on demand and stored for reuse. They do not modify the underlying dataset.

\subsection{Spectrum Viewer}

The Spectrum Viewer displays the dataset's channels on the horizontal axis and intensity on the vertical axis. For each segment, it shows a configurable summary spectrum (minimum, maximum, or average) over all pixels belonging to that segment, color-coded by segment color. The summary is computed channel-wise, i.e., the statistic is taken independently per channel across the segment's pixels, yielding a spectrum rather than a single scalar value. The global summary over all pixels is shown in black as a reference.

Users interact with the Spectrum Viewer primarily to create features and to preprocess the dataset. Clicking or dragging over the spectral axis creates a new feature by selecting a single channel or integrating a range; integration supports multiple accumulation modes and baseline correction options (from zero, from the local minimum, or from a linear baseline connecting the range endpoints). Dataset-wide preprocessing operations include baseline correction and the computation of spectral derivatives up to the third order, which are applied in-place and can substantially improve cluster separation in the embedding, as demonstrated in Section~\ref{sec:uc1}. Since a spectrum is sampled at discrete channels, the derivative is computed as a finite difference between neighboring channels along the (regularly sampled) spectral axis. It is therefore a discrete approximation defined between adjacent channels rather than a derivative of a continuous function.

The Spectrum Viewer also supports spectral similarity analysis: users can import one or more reference spectra and compute a pixelwise similarity image using either Euclidean distance or cosine similarity between each pixel's spectrum and the reference spectrum, with both treated as vectors over the shared channels. The resulting similarity values are exported as a new dataset, enabling downstream spatial analysis. Reference spectra can additionally be exported for use in other tools or shared across sessions.

\subsection{Image Viewer}

The Image Viewer provides spatial context by rendering pixels at their measured $x$/$y$ coordinates, mapping the values of a selected feature to colors via a configurable colormap. A segmentation overlay or false-coloring can be displayed at adjustable opacity, allowing the feature image and the segmentation to be inspected simultaneously using $\alpha$-blending.

The segmentation can be modified interactively in the Image Viewer using lasso selections. Automatic segmentations can be created using $K$-means, HDBSCAN~\cite{mcinnes_2017_hdbscan}, or the Leiden algorithm~\cite{traag_2019_leiden}, applied either to the raw dataset or to a previously computed embedding. Segments can be added, removed, renamed, merged, and recolored through the context menu, and segmentations can be imported and exported for reproducibility.

\begin{figure}[tb]
    \centering
    \includegraphics[width=\linewidth]{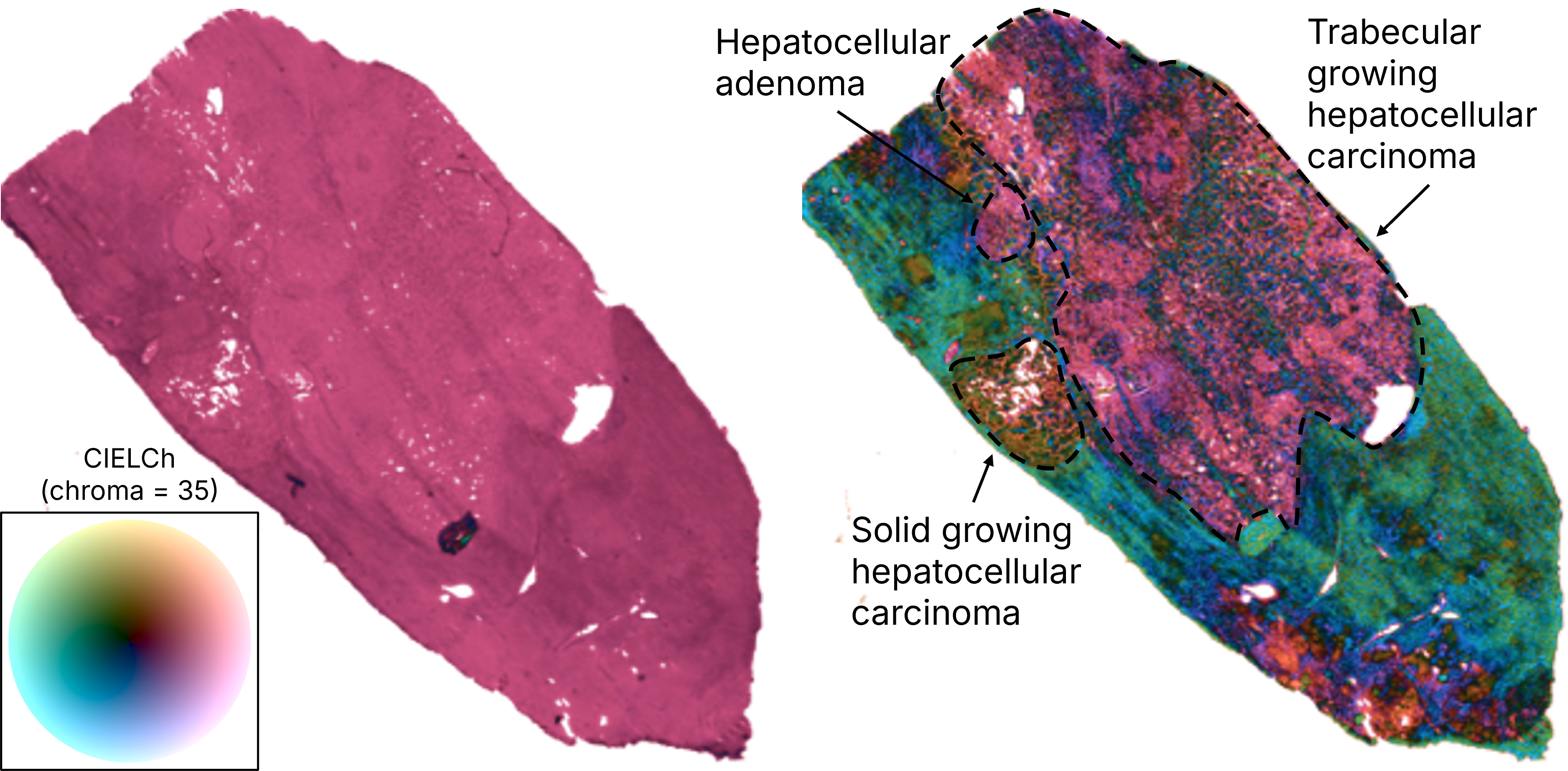}
    \caption{False-coloring of a 213-channel IR microscopy image of a tumor-bearing rat liver~\cite{kronenberg_2023_multimodal}, derived from a PCA embedding of the pixel spectra. Without contrast enhancement (left), dense regions of the embedding compress into similar colors. With the density-equalizing transformation applied (right), tissue substructures (indicated by manual annotations) become more visible. The color encodes each pixel's position in the embedding, mapped onto the fixed-chroma CIELCh color disk shown as an inset (bottom left).}
    \label{fig:false_coloring}
\end{figure}

False-coloring assigns colors to pixels based on their position in a 2D embedding rather than on segment membership, providing a continuous coloring that reveals fine-grained spatial structure even within segments. To increase contrast, MIA applies a density-equalizing transformation~\cite{rave_2026_multiresolution} to the embedding coordinates before mapping them into polar coordinates and into a fixed-chroma (chroma $=\num{35}$) disk of the CIELCh color space, increasing inter-point distances in dense regions while preserving neighborhood structure, see Figure~\ref{fig:false_coloring}.

\subsection{Embedding Viewer}

The Embedding Viewer displays a 2D embedding of a selected set of channels or features for a chosen subset of pixels, with each point representing one pixel. Points are colored by segment membership or by false-coloring, and the view is linked to the Image Viewer such that lasso selections in the embedding space are immediately reflected as spatial selections in the image.

Embeddings are computed using UMAP~\cite{mcinnes_2018_umap}, PCA~\cite{jolliffe_2005_principal}, or t-SNE~\cite{maaten_2008_tsne}, with user control over which pixels, channels, and features are included, the normalization strategy (z-score, min-max, or none), and whether to export the underlying model for projection of new data.

For large datasets, MIA computes the UMAP embedding on a representative subset of pixels and projects the remaining pixels onto the learned embedding using UMAP's out-of-sample transform~\cite{mcinnes_2018_umap}. We refer to this subsample-and-project scheme as \textit{landmark-based UMAP}. It is an approximation built on standard UMAP rather than a distinct embedding method. The subset is drawn by uniform random sampling of pixels. This makes interactive exploration of datasets with millions of pixels feasible without requiring specialized hardware~(G4).

\textit{Hierarchical embeddings} allow users to iteratively refine the analysis by re-embedding only the pixels belonging to one or more selected segments. This is particularly useful if a global embedding reveals broad macro-structure but obscures finer distinctions within individual clusters; re-embedding a subcluster at higher effective resolution can reveal meaningful sub-populations, as demonstrated in Section~\ref{sec:uc1}. Pixels not yet assigned to a segment can be assigned to their nearest neighbor in the embedding space, supporting a semi-automatic segmentation workflow.

\subsection{Channel Glyph Viewer}

\begin{figure}[tb]
    \centering
    \includegraphics[width=0.95\linewidth]{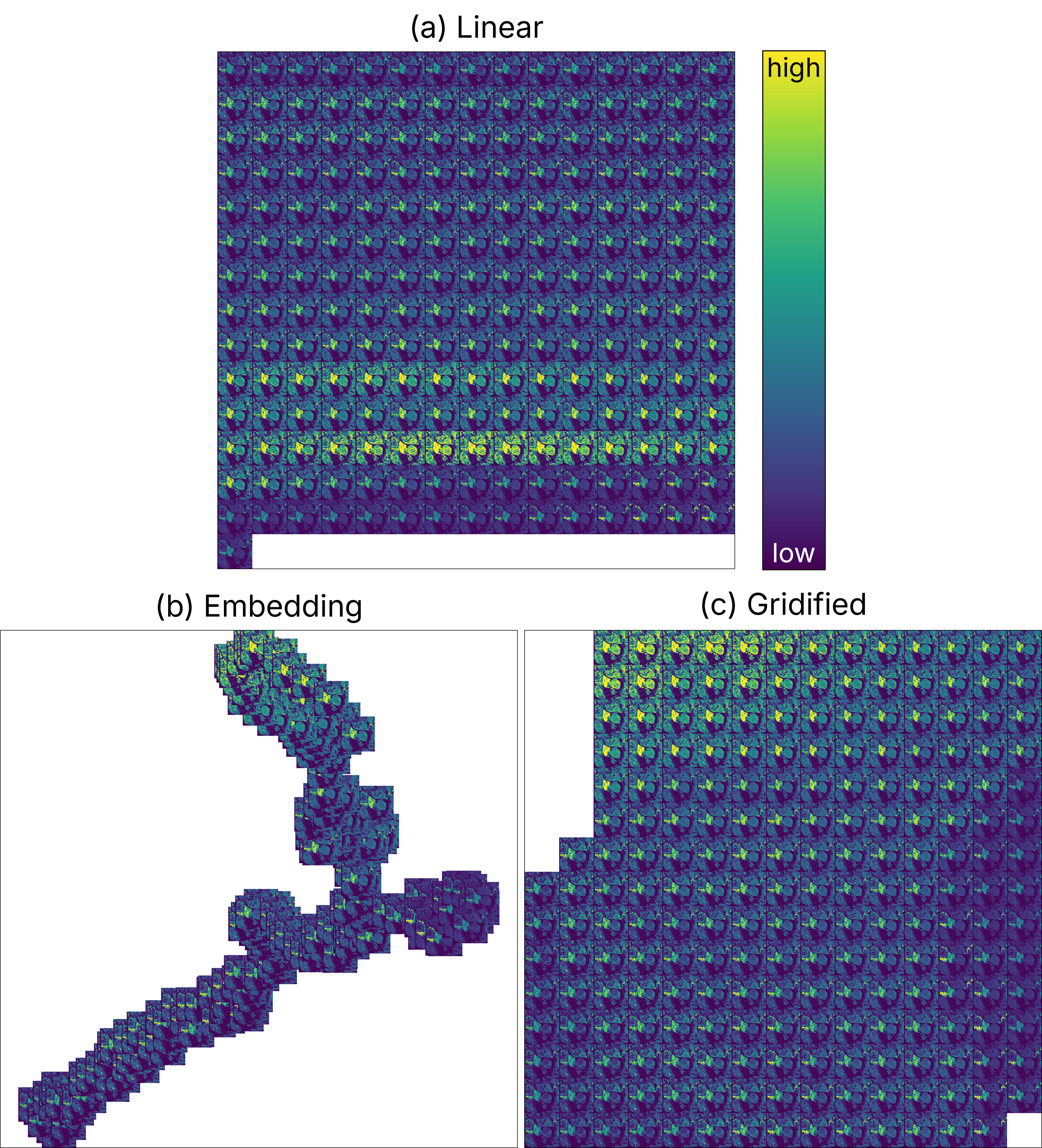}
    \caption{The three layout modes of the Channel Glyph Viewer, shown on a QCL-IR microscopy dataset with 211 spectral channels. (a)~\textit{Linear} mode arranges glyphs row by row in channel order. (b)~\textit{Embedding} mode positions glyphs according to a UMAP embedding of the channels. (c)~\textit{Gridified} mode resolves the resulting overlaps of the embedding by moving glyphs to nearby discrete grid positions via Linear Sum Assignment.}
    \label{fig:channel_embedding}
\end{figure}

The Channel Glyph Viewer provides a glyph-based overview of the entire dataset at the level of channels rather than pixels. Each channel is represented as a small glyph -- a miniature image of that channel's values rendered over the pixels' spatial coordinates, i.e., a thumbnail of the channel's spatial distribution -- with a customizable color map. Such glyph-based designs compactly encode rich per-item information within a spatial layout~\cite{stoppel_2016_graxels}. The layout of the glyphs can be configured in three modes, see Figure~\ref{fig:channel_embedding}. In \textit{linear} mode, glyphs are arranged row by row in channel order, providing a compact overview of all channels simultaneously. In \textit{embedding} mode, glyph positions are determined by a UMAP embedding of the channels: here, each channel is treated as a data point whose similarity to every other channel is computed across the pixels (equivalently, on the transposed data matrix) using a selected similarity metric (Pearson correlation, mutual information, Euclidean distance, or cosine distance), thus, effectively grouping similar channels together. In \textit{gridified} mode, overlaps between glyphs are resolved by assigning them to discrete grid positions using Linear Sum Assignment~\cite{burkard_2009_assignment}, which computes the one-to-one assignment of glyphs to grid cells that minimizes the total displacement from their embedding positions, thus preserving the neighborhood structure of the embedding while ensuring legibility.

Per-segment channel abundances, computed by accumulating intensities per channel per segment, can be overlaid as donut charts, providing a concise summary of how intensity is distributed across segments for every channel simultaneously (see supplementary video).

In practice, users have also employed the Channel Glyph Viewer in linear mode as a spatial overview of similarity images computed in the Spectrum Viewer, treating each channel as one similarity map. This emergent use was not part of the original design intent, and is discussed further in Section~\ref{sec:uc2}.

\subsection{Histogram and Boxplot Viewers}

The Histogram Viewer displays the distribution of intensities for a selected feature as a stacked histogram, with each segment contributing a separate colored layer. The Boxplot Viewer shows the same information as individual boxplots for each segment alongside a boxplot for the dataset as a whole, supporting rapid visual comparison of per-segment intensity ranges and outlier structure. Both viewers respond to changes in the active segmentation and feature selection. These views are used less frequently than the embedding and image views in typical workflows, but provide complementary distributional information per segment.

\subsection{Multimodal Analysis}
\label{sec:multimodal}

MIA supports the simultaneous analysis of multiple co-registered datasets from different modalities. Having loaded a dataset, an additional dataset is imported via the Spectrum Viewer and, if the two datasets differ in spatial resolution, a registration dialog allows the user to manually align them. The imported dataset is then resampled to match the spatial resolution of the primary dataset using a user-selected interpolation scheme (nearest neighbor or bilinear). Each imported dataset is opened in a separate, independently resizable application window containing the full set of coordinated views, allowing each modality to be analyzed individually at full size. The windows are not sub-panels of a fixed layout, so feature legibility is not compromised. The segmentation is shared across all windows, ensuring that every segment is defined consistently, regardless of how many co-registered modalities are loaded.

Multimodal embeddings are created in the Embedding Viewer by selecting channels or features from multiple datasets simultaneously. When more than one modality is selected, a weighting dialog allows the user to control the relative contribution of each modality to the final embedding. This is important in practice because modalities can differ substantially in channel count, and naive concatenation would cause the high-channel-count modality to dominate the embedding geometry.

\section{Use Cases}
\label{sec:usecases}

\begin{figure*}[!htb]
    \centering
    \includegraphics[width=\linewidth]{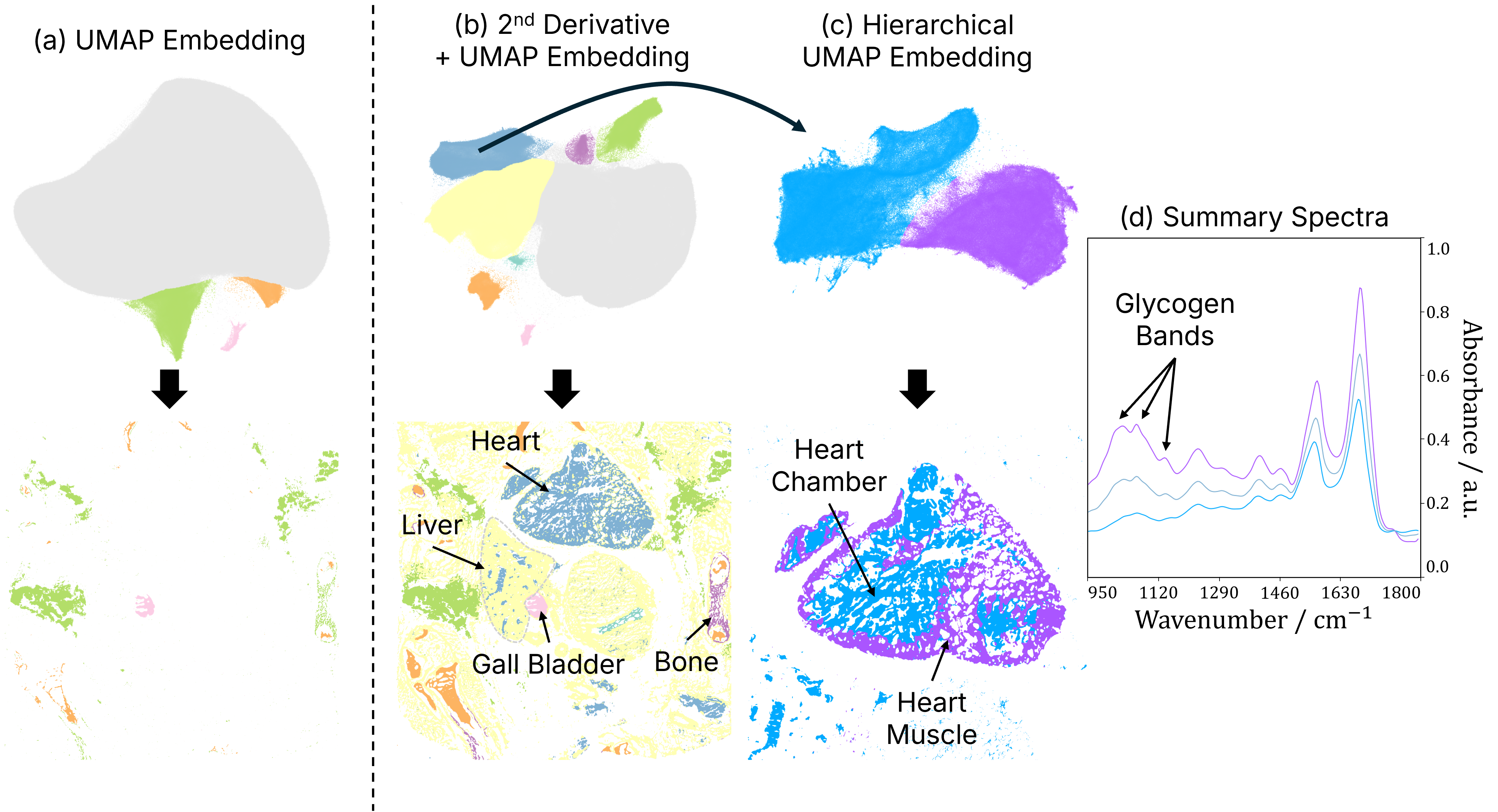}
    \caption{Use Case~1: Tissue segmentation via hierarchical embedding on a QCL-IR microscopy dataset of a chicken embryo (tissue cross-section of the torso).
        (a)~UMAP embedding (top) and image view (bottom) computed on raw spectra: no meaningful cluster structure is apparent.
        (b)~After computing the second derivative along the spectral dimension, the UMAP embedding reveals clear cluster structure corresponding to spatially coherent tissue regions.
        (c)~Interactive re-embedding of a single subcluster (i.e., computing a new UMAP on the selected pixels) separates the heart chamber from the heart muscle.
        (d)~Spectrum Viewer showing the average spectra of the two subclusters compared to the average spectrum of the parent cluster before hierarchical refinement: the subclusters are chemically distinct, while the parent spectrum is a blurred average of both signatures, confirming that the hierarchical embedding successfully resolves two tissue types that the global embedding grouped together.}
    \label{fig:uc1}
\end{figure*}

\begin{figure*}[!htb]
    \centering
    \includegraphics[width=\linewidth]{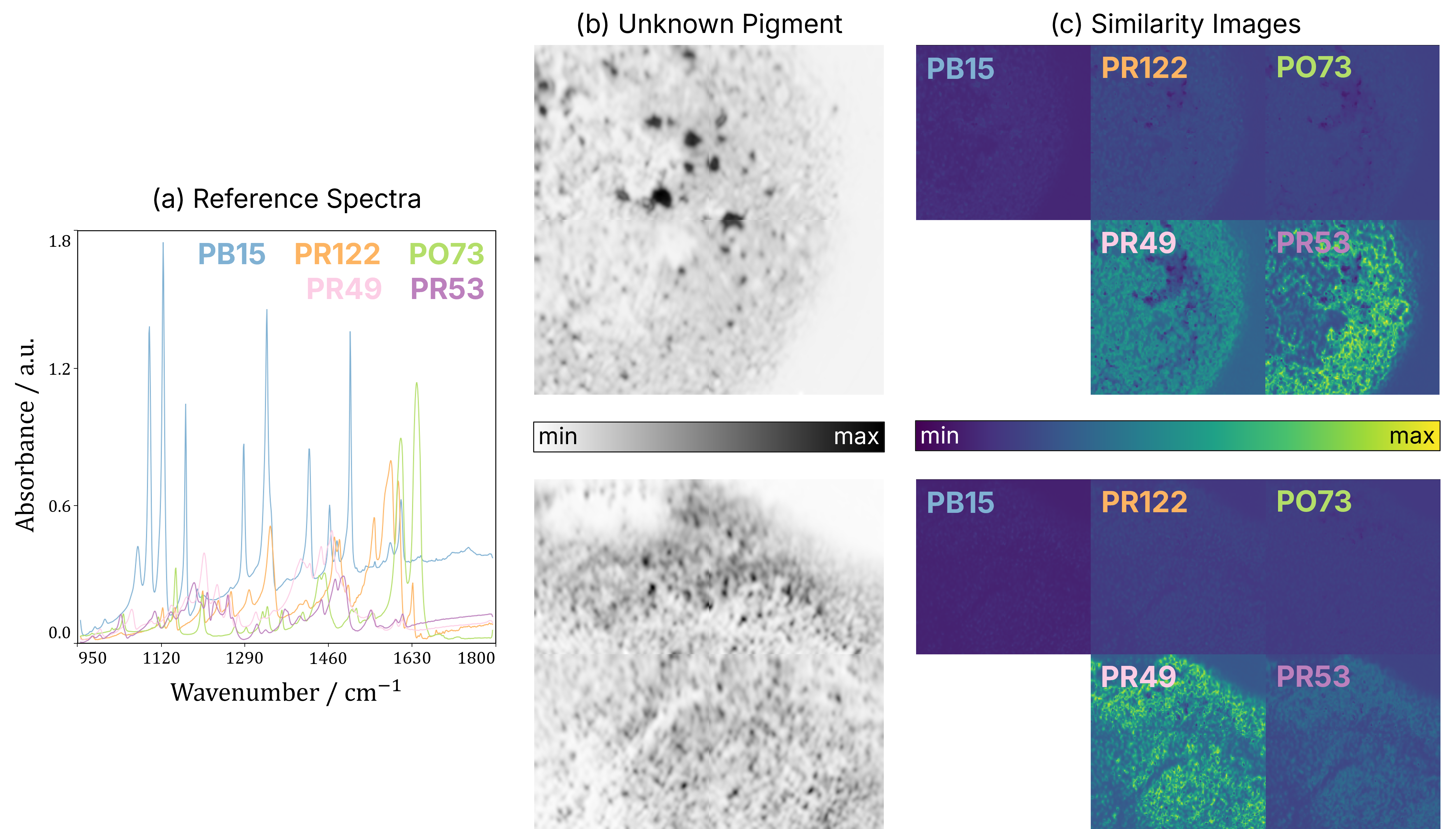}
    \caption{Use Case~2: Spectral similarity search on two QCL-IR microscopy datasets of pigment samples (PR53 and PR49). (a)~The five reference spectra representing known pigments, shown in the Spectrum Viewer. (b)~Grayscale images of the two datasets showing the spatial distribution of the analyzed pigment on the slide. (c)~Channel Glyph Viewer in linear mode, displaying the five similarity images for each dataset as glyphs: PR53 shows the highest similarity to the PR53 reference (top), and PR49 to the PR49 reference (bottom), while the structural similarity between the two pigments produces moderate cross-matches in both cases.}
    \label{fig:uc2}
\end{figure*}

\begin{figure*}[!htb]
    \centering
    \includegraphics[width=\linewidth]{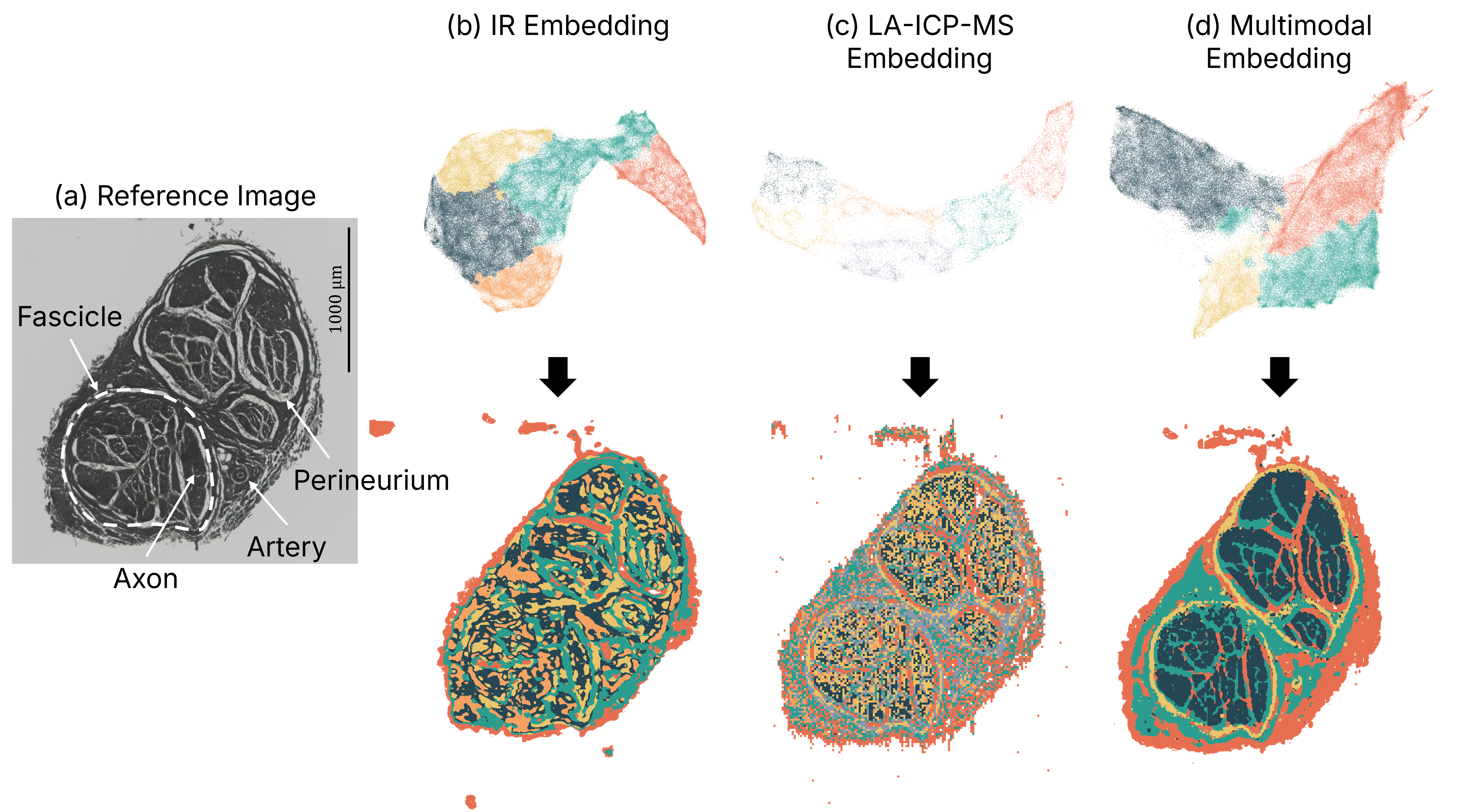}
    \caption{Use Case~3: Multimodal embedding of co-registered QCL-IR microscopy and LA-ICP-MS data from a transverse section of human sciatic nerve tissue. (a)~Brightfield reference image showing the anatomical structure of the section.
        (b)~UMAP embedding and segmentation derived from IR data alone, resolving molecular tissue compartments.
        (c)~UMAP embedding and segmentation derived from LA-ICP-MS data alone, reflecting elemental composition.
        (d)~Combined multimodal embedding and segmentation: the integration of both modalities yields clearer compartment separation; the gadolinium signal is concentrated in a single, chemically distinct cluster, shown here in yellow-beige.}
    \label{fig:uc3}
\end{figure*}

The following use cases are drawn from real analytical workflows developed in close collaboration with domain experts in analytical chemistry. In each case, the dataset and workflow were provided by a collaborating researcher. We reproduce the analysis steps within MIA and report the results. The use cases are ordered from the most common workflow pattern to more specialized applications, reflecting the range of scenarios encountered in practice. Together, they illustrate the design goals described in Section~\ref{sec:goals}: the integrated workflow~(G1), shared segmentation state~(G2), and iterative exploration~(G5) are central to all three cases; scalability~(G4) is demonstrated in Use Case~1; and modality-agnosticism~(G3) is central to Use Case~3.

\subsection{Use Case 1: Tissue Segmentation via Hierarchical Embedding}
\label{sec:uc1}

\paragraph{Dataset and goal}
The first use case is based on a QCL-IR microscopy dataset of a tissue cross-section of a chicken embryo torso. The dataset consists of \num{9000000}~pixels at a spatial resolution of \num{3000} $\times$ \num{3000} and \num{211}~spectral channels covering the mid-infrared fingerprint region from \num{950} to \num{1800}~cm$^{-1}$. The analytical goal is to segment the tissue into biologically meaningful regions, in particular, to separate distinct tissue types that differ in chemical composition, and to characterize their spectral signatures.

\paragraph{Step 1: Embedding of raw spectra}
As a first step, a UMAP embedding is computed on the raw spectra using all \num{211}~channels. Figure~\ref{fig:uc1}a shows the resulting embedding alongside the corresponding image view. Besides one small detached cluster corresponding to the gall bladder, the embedding reveals only a broad, poorly separated structure: a single dominant group of points with two minor protrusions, which are difficult to delineate cleanly into interpretable tissue regions. This result is typical for raw IR spectra, where baseline variation and broad, overlapping absorption bands dominate the inter-pixel distances and obscure the underlying chemical differences.

\paragraph{Step 2: Spectral derivative preprocessing}
To suppress baseline effects and enhance the discriminability of spectral features, the second derivative is computed along the spectral dimension across the entire dataset using the Spectrum Viewer. The second derivative attenuates broad baseline contributions and resolves shoulder bands within overlapping absorption features, effectively acting as a high-pass filter in the spectral domain. A new UMAP embedding is then computed on the derivative spectra. As shown in Figure~\ref{fig:uc1}b, the resulting embedding exhibits substantially clearer cluster structure: the previously amorphous group of points resolves into several distinct clusters corresponding to the heart, bone, and a larger cluster encompassing the liver and surrounding tissue, while the gall bladder remains a well-defined, isolated cluster as before.

\paragraph{Step 3: Hierarchical embedding for subcluster refinement}
While the second-derivative embedding provides a useful initial segmentation, the heart cluster contains pixels from two spatially distinct regions that are not resolved at the global embedding scale. To refine the analysis, a hierarchical embedding is computed: only the pixels belonging to this cluster are selected, and a new UMAP embedding is computed on this subset. This increases the effective resolution of the embedding within the subcluster, allowing finer spectral differences to drive the point layout. Figure~\ref{fig:uc1}c shows that the hierarchical embedding clearly separates the subcluster into two distinct groups. Projecting this subdivision back onto the image view reveals that the two groups correspond to the \textit{heart chamber} and the \textit{heart muscle}, two tissue types that are histologically distinct but spectrally similar enough to appear as a single cluster in the global embedding.

\paragraph{Step 4: Spectral validation}
To validate the segmentation, Figure~\ref{fig:uc1}d compares the per-segment average spectra of the two subclusters against the average spectrum of the parent cluster from which they were derived -- i.e., the single cluster containing the entire heart region prior to hierarchical refinement. The spectra of the two embedding-derived subclusters are clearly distinct, with the heart muscle exhibiting substantially higher absorbance in the glycogen band region when compared to the heart chamber. By contrast, the parent cluster spectrum is a blurred average of both signatures, reflecting the fact that the initial global embedding grouped these two chemically distinct tissue types together. This comparison demonstrates that the hierarchical embedding recovers chemically pure tissue signatures that the global embedding alone cannot resolve.

\subsection{Use Case 2: Spectral Similarity Search and Spatial Overview}
\label{sec:uc2}

\paragraph{Dataset and goal}
The second use case is based on two QCL-IR microscopy datasets of small areas of pigment samples deposited on an indium tin oxide-coated glass slide. Each dataset consists of \num{250000}~pixels across \num{843}~channels. The analytical goal is to identify which of five known reference pigment spectra best matches the sample, supporting visual identification through spatial distribution and providing evidence for structural group membership based on spectral similarity patterns. The five reference pigments are Pigment Red 53:1~(PR53), Pigment Red 49~(PR49), Pigment Blue 15~(PB15), Pigment Orange 73~(PO73), and Pigment Red 122~(PR122). This workflow is representative of pigment authentication and verification tasks, where a measured sample must be matched against a library of known references.

\paragraph{Computing similarity images}
The five reference spectra are imported into the Spectrum Viewer. Before computing similarities, a baseline correction is applied to the dataset to reduce the influence of broad spectral offsets on the distance metric. Pixelwise similarities are then computed against each reference spectrum using a normalized Euclidean distance metric $s = 1/(1+d)$, where $d$ is the Euclidean distance between the pixel spectrum and the reference, yielding similarity scores $s \in (0, 1]$ with higher values indicating a closer match. For each reference spectrum, MIA produces one similarity image: a spatially resolved map in which intensity encodes the degree of spectral match at each pixel. The resulting similarity images are exported as a new dataset, effectively treating each reference spectrum as a channel in a derived dataset. Figure~\ref{fig:uc2}b shows the spatial distribution of the two pigment samples, confirming that the material is distributed across the slide in a spatially coherent pattern that can serve as a basis for similarity analysis.

\paragraph{Spatial overview via the Channel Glyph Viewer}
With five similarity images per dataset, visual inspection of individual maps is straightforward, but comparing patterns across all references simultaneously benefits from an overview. The derived similarity datasets are loaded into the Channel Glyph Viewer in \textit{linear} mode, where each channel -- here, each similarity image -- is displayed as a small glyph arranged row by row. Figure~\ref{fig:uc2}c shows the result for both datasets: in each case, the highest similarity is observed for the matching reference pigment -- PR53 for the first dataset and PR49 for the second -- revealing the identity of the samples. Because PR53 and PR49 are structurally related pigments, moderate cross-similarity is also visible in both datasets, which is chemically expected and itself informative about the structural group membership of the sample. The Channel Glyph Viewer makes this pattern immediately apparent without requiring the analyst to inspect each similarity image individually. The best-matching reference could in principle be determined automatically, by comparing aggregate similarity scores across references. The visual overview adds what such a score cannot: it confirms that a match is spatially coherent across the sample rather than driven by a few high-similarity pixels or localized artifacts, it exposes the full cross-similarity pattern that signals structural-group membership (here between PR53 and PR49), and it would reveal spatial heterogeneity or the presence of multiple components that a single aggregate score would obscure.

\paragraph{Emergent use}
This use case illustrates an emergent pattern not anticipated in the original design of the Channel Glyph Viewer, which was conceived primarily as a tool for exploring the full channel space of a hyperspectral dataset. In practice, users discovered that the linear layout serves equally well as a spatial overview of derived feature sets such as similarity images, repurposing a general-purpose view for a specific analytical task. This reflects a broader design principle of MIA: by keeping views general and linked through a shared data model, the tool accommodates analytical strategies that were not explicitly designed for.

\subsection{Use Case 3: Multimodal Embedding}
\label{sec:uc3}

\paragraph{Dataset and goal}
The third use case demonstrates MIA's multimodal analysis capabilities using a co-registered dataset combining QCL-IR microscopy and LA-ICP-MS measurements acquired on a parallel tissue section. The sample is a transverse thin section of human peripheral nerve tissue obtained from an autopsy. The scientific question motivating this dataset is whether gadolinium, a metallic element administered as a contrast agent during MRI examinations, is retained in the tissue after the procedure and, if so, how its distribution relates to the molecular tissue composition. This question requires both molecular tissue characterization (provided by IR microscopy across \num{211}~spectral channels at \SI{5}{\micro\metre} pixel size) and spatially resolved elemental composition (provided by LA-ICP-MS across \num{4}~channels (Gd, Fe, Zn, P) at \SI{15}{\micro\metre} pixel size), making it a natural candidate for multimodal analysis.

\paragraph{Individual embeddings}
For the IR dataset, baseline correction and second-derivative preprocessing are applied before computing a UMAP embedding, followed by Leiden clustering. Figure~\ref{fig:uc3}b shows the resulting embedding and segmentation: the IR data resolves the major molecular tissue compartments of the peripheral nerve, including fascicles, perineurium, interfascicular connective tissue, and vascular structures, in good agreement with the brightfield reference image shown in Figure~\ref{fig:uc3}a. Figure~\ref{fig:uc3}c shows the analogous result for the LA-ICP-MS data alone, using the same embedding parameters applied to the four elemental channels: the elemental data captures a different aspect of tissue composition, reflecting the distribution of metals and phosphorus rather than molecular structure. While both individual embeddings produce anatomically interpretable segmentations, each reflects only one dimension of the tissue's chemical composition.

\paragraph{Multimodal embedding}
The two datasets are loaded simultaneously into MIA, which resamples the LA-ICP-MS data to the IR spatial resolution (\SI{5}{\micro\metre}) using bilinear interpolation. A combined embedding is then computed using channels from both modalities. Because the IR dataset contains \num{211}~channels compared to \num{4}~elemental channels, the modality weighting dialog is used to balance their contributions to the embedding geometry; without weighting, the IR modality would dominate, and the elemental information would have negligible influence on the result. In this analysis, the modalities were weighted 1:2 (IR:LA-ICP-MS), giving the elemental information a stronger influence on the embedding. Figure~\ref{fig:uc3}d shows the resulting multimodal embedding and segmentation: the integration of both modalities yields clearer separation of anatomical compartments and more stable, consistent clusters than either modality alone. Crucially, the multimodal segmentation enables direct spatial correspondence between gadolinium-containing regions identified by LA-ICP-MS and the specific molecular tissue compartments characterized by IR. In this dataset, the gadolinium signal is concentrated in a single cluster, which appears as the yellow-beige segment in the multimodal segmentation~(d), as confirmed by a domain expert inspecting the per-segment spectra. This demonstrates how the multimodal embedding localizes an elemental signal within a molecularly defined region that neither modality resolves on its own.


\section{Discussion}
\label{sec:discussion}

\subsection{Expert Feedback}

MIA has been developed in close collaboration with researchers in analytical chemistry and has been used in published research involving IR microscopy, micro-XRF, and LA-ICP-MS data~\cite{kronenberg_2023_multimodal, peivandi_2025_cardiovascular, kronenberg_2025_umap}.
To systematically gather feedback, we distributed a written questionnaire including the System Usability Scale (SUS)~\cite{brooke_1996_sus} to \num{6}~domain experts. We note that the respondents are also collaborators on the development of MIA, which limits the independence of the evaluation. MIA received an average SUS score of \num{84.2} (SD = \num{8.76}), corresponding to a rating of ``good'' on the Bangor et al.\ adjective scale~\cite{bangor_2009_determining}. The individual responses and the full questionnaire are provided in the supplementary material. The following discussion summarizes the qualitative findings. We deliberately evaluate MIA through expert feedback and real-world use cases rather than a controlled comparative experiment. This follows the established methodology for design studies and domain tool papers~\cite{sedlmair_2012_design_study,lam_2012_seven}, in which qualitative, case-based validation is the appropriate scenario: the contribution is an integrated environment validated against authentic analytical workflows, not an isolated technique whose performance can be measured against a baseline in a controlled task. A controlled comparison against existing tools would also be confounded here, as the respondents are co-developers; we therefore report this evaluation transparently as expert feedback and discuss its limitations explicitly.

Collaborators consistently identified the combination of UMAP-based embedding and interactive segmentation as the most valuable part of MIA's workflow; all six respondents selected embedding as a frequently used feature, and manual as well as automatic segmentations were each selected by five of six respondents (four reported regularly using both). The ability to compute embeddings hierarchically within subclusters was specifically highlighted as enabling a level of analytical refinement that was not achievable with prior tools. As illustrated in Use Case~1 in Section~\ref{sec:uc1}, this iterative process revealed tissue-level distinctions that were invisible in the raw data and that the global embedding alone could not resolve. One respondent noted that ``MIA has made my workflow more iterative and exploratory. I spend less time on manual preprocessing and switching between tools.'' Another observed that ``segmentation based on dimensionality reduction helps to quickly identify substructures and patterns in the data.'' Asked specifically whether MIA enabled discoveries that were harder to obtain with their previous tools, respondents pointed to concrete examples: correlating LA-ICP-MS results with histological tissue features (cf.\ Use Case~3 in Section~\ref{sec:uc3}), screening and managing the full set of $m/z$ channels of ICP-TOFMS data prior to export, and identifying patterns and relationships that they had not explicitly set out to look for. These responses substantiate the claim that the integrated, embedding-driven workflow surfaces structure that is difficult to reach with the disconnected tools used previously.

The reduction in tool-switching was a recurring theme across respondents. When asked about their prior workflows, respondents reported using a mix of commercial vendor software, self-coded Python scripts, or ad-hoc combinations of individual analysis packages. MIA consolidates these steps into a single interface with a shared segmentation state, and respondents noted that this not only reduced time but also reduced the risk of inconsistencies introduced by transferring data between tools. Notably, this group included respondents who had previously written their own Python analysis scripts, indicating that the benefit of integration is not confined to users without a programming background.

The emergent use of the Channel Glyph Viewer for spatial overview of similarity images (see Use Case~2 in Section~\ref{sec:uc2}) was not identified by us but discovered independently by collaborators during routine use. This finding reflects the value of providing flexible, general-purpose views that users can adapt to analytical needs beyond those explicitly anticipated in the design.

\subsection{Comparison to Scripting-Based Workflows}
A natural question is how MIA compares to a general-purpose scripting environment, such as a Jupyter notebook built on the same underlying libraries (UMAP, scikit-learn). While we do not claim a controlled comparison, several differences are decisive for the target users and tasks. (1) Accessibility: as reflected in the questionnaire, the intended users are largely not programmers, and a notebook-based workflow presupposes coding proficiency that they, generally, do not have. (2) Performance: MIA is implemented in C++ and renders and re-segments datasets of millions of pixels interactively, whereas equivalent operations in an interpreted notebook are typically too slow for the fluid, iterative exploration that MIA is designed around. (3) Interactivity: the shared segmentation state propagates every lasso selection, clustering, or edit across all linked views instantaneously, providing immediate visual feedback that a cell-by-cell notebook, which re-renders static figures on demand, cannot match. (4) Reproducibility and consistency: segmentations, embeddings, and reference spectra can be exported and re-imported, and the single shared state removes the manual data transfers between tools that respondents identified as a source of inconsistency. Conversely, scripting environments retain advantages in flexibility and extensibility, and we view the two as complementary rather than mutually exclusive. Tellingly, the questionnaire indicates that even respondents who had previously written their own Python analysis scripts adopted MIA for these exploratory tasks.

\subsection{Limitations and Future Work}

The following directions for future development were identified through collaborator feedback and reflect the boundaries of MIA's current scope.


\paragraph{Embedding alignment for consistent false-coloring}
When comparing multiple embeddings of the same dataset -- for example, embeddings computed with different parameters or applied to different modalities of a co-registered multimodal dataset -- the false-coloring assigned to each embedding is independent, as the color mapping depends on the absolute positions of points in the 2D embedding space. This makes visual comparison across embeddings difficult, since corresponding pixels may receive entirely different colors in each view. A natural extension would be to align embeddings that share a
point-to-point correspondence via a rigid transformation with uniform scaling, bringing their coordinate frames into a common reference and ensuring that corresponding pixels receive consistent colors across embeddings. This would be particularly valuable for the multimodal use case described in Section~\ref{sec:uc3}, where comparing per-modality and combined embeddings side by side is a central analytical step.


\paragraph{Dedicated hierarchical embedding}
    MIA's hierarchical workflow relies on interactive, user-driven re-embedding of selected segments rather than a dedicated hierarchical dimensionality reduction algorithm. While this integrates tightly with the shared segmentation state and spectral preprocessing, it does not construct an explicit multiscale hierarchy as methods such as HSNE~\cite{https://doi.org/10.1111/cgf.12878}, HUMAP~\cite{marcilio_2024_humap}, or Multiscale PHATE~\cite{kuchroo_2022_multiscale_phate} do. These methods precompute a hierarchy that can be interactively sliced to re-embed a user-selected subset on the fly. Incorporating such a method as an additional embedding backend would complement MIA's manual refinement with an automatically derived multiscale structure, allowing users to navigate levels of granularity without recomputing embeddings, and represents a promising direction for future development.

\paragraph{Volumetric data}
    MIA's data model associates each pixel with a pair of spatial coordinates, i.e., it assumes a 2D imaging geometry. A number of modalities (for example, serial-section or depth-resolved acquisitions) instead produce volumetric data, in which each measurement is a voxel with three spatial coordinates. Extending MIA to voxels would require adapting the Image Viewer to volume rendering and slice-based navigation, while the spectral, embedding, and segmentation machinery would remain largely unchanged, since they already operate on the spatial arrangement only through the coordinates attached to each measurement. We regard this as a natural and largely incremental extension.

\paragraph{MALDI imaging with ion mobility spectrometry}
Matrix-assisted laser desorption/ionization (MALDI) imaging with ion mobility separation produces data with a fundamentally different structure: each pixel is associated with a two-dimensional spectrum indexed jointly by mass-to-charge ratio ($m/z$) and reduced inverse ion mobility ($1/K_0$), rather than the one-dimensional spectrum along a single spectral axis assumed by MIA's data model. Supporting this modality would require extending the data model to accommodate 2D spectra per pixel, with corresponding changes to the Spectrum Viewer and feature computation. Given the growing importance of ion mobility spectrometry in molecular MSI, this represents a significant and worthwhile direction for future development.

\paragraph{Cohort analysis}
MIA's multimodal framework currently requires datasets to be spatially co-registered, as the shared segmentation state is defined over a common set of pixels. This precludes a related but distinct analytical scenario: cohort analysis, in which multiple datasets from the same modality are acquired from different specimens or patients, and the analyst wishes to compute a combined embedding across the entire cohort without spatial alignment. This capability -- loading two or more datasets of the same modality for a joint embedding -- was independently requested by a collaborator in the questionnaire, confirming its practical relevance. Supporting cohort analysis would require a fundamentally different segmentation model and would open MIA to a broad class of comparative studies in biomedical imaging. We consider this the most consequential direction for future work.

\paragraph{Applicability to other hyperspectral domains}
MIA's data model makes no domain-specific assumptions, so its workflow extends in principle to hyperspectral imaging beyond analytical chemistry, such as remote sensing or agriculture. Realizing this potential would require accommodating the differing acquisition conventions and analytical questions of these domains, which we see as a promising avenue for broadening MIA's reach.

\section{Conclusion}
\label{sec:conclusion}

We presented MIA, a modality-agnostic visual analysis environment for hyperspectral bioimaging that integrates spectral preprocessing, embedding-based dimensionality reduction, interactive segmentation, and spectral exploration within a single linked-view interface. By maintaining a shared segmentation state across all views and supporting a common data model for IR microscopy, LA-ICP-MS, and their multimodal combinations, MIA reduces the tool-switching overhead that characterizes current analytical workflows in this domain.

Three use cases drawn from real collaborator workflows demonstrate the breadth of analytical scenarios MIA supports. The first use case showed that interactive re-embedding of subclusters applied to derivative-preprocessed IR spectra recovers biologically meaningful tissue distinctions (heart chamber versus heart muscle) that are invisible in raw spectral data and unattainable through manual segmentation alone. The second use case demonstrated that spectral similarity analysis, combined with the Channel Glyph Viewer repurposed as a spatial overview tool, enables efficient identification of regions matching known reference compounds. The third use case illustrated how combining IR and LA-ICP-MS data in a weighted multimodal embedding yields clearer separation of anatomical tissue compartments and enables direct localization of the gadolinium signal within specific molecular tissue environments -- an insight not accessible from either modality individually. Together, these results demonstrate that integration and interactivity are independent contributions from algorithmic novelty in the development of analytical software for the biomedical and chemical imaging sciences. We hope that MIA lowers the barrier to sophisticated exploratory analysis for domain experts who are not computational specialists, and that its open design provides a foundation for future extensions to emerging modalities and study designs.

\section*{Acknowledgments}

This work was supported in part by Deutsche Forschungsgemeinschaft (DFG) -- CRC 1450 -- 431460824. The authors thank the Stiftung der Deutschen Wirtschaft gGmbH (sdw) for a scholarship to Hannes Gödde. The presented results have been achieved in the framework of the EFRE.NRW-funded project Pig-Pro-QuO (EFRE-20800668; IN-WP-2-005). We thank Astrid Jeibmann, Clinic for Diagnostic Imaging, Diagnostic Imaging Research Unit (DIRU), University of Zürich, for her support. We thank Dr. Nassim Ghaffari-Tabrizi-Wizsy from the Otto Loewi Research Center, Division of Immunology, Medical University of Graz, for her valuable support with the chicken embryo assay.

\bibliographystyle{cag-num-names}
\bibliography{refs}



\end{document}